\documentclass[12pt,a4paper]{article}
\usepackage[utf8]{inputenc}
\usepackage[T1]{fontenc}
\usepackage{amsmath,amssymb,bbm}
\usepackage{graphicx}
\usepackage{color}

\begin{document}

\textwidth=135mm
 \textheight=200mm
\begin{center}
{\bfseries Dynamics and gravitational-wave emission of neutron-star merger remnants}
\vskip 5mm
A. Bauswein$^{\$,}$$^*$ and J. Clark$^{\dag}$ and N. Stergioulas$^*$ and H.-T. Janka$^{+}$
\vskip 5mm
{\small {\it $^\$$ Heidelberger Institut f\"ur Theoretische Studien, Schloss-Wolfsbrunnenweg~35, D-69118~Heidelberg, Germany}}\\
{\small {\it $^*$ Department of Physics, Aristotle University of Thessaloniki, GR-54124 Thessaloniki, Greece}}\\
{\small {\it $\dag$ Georgia Institute of Technology, Atlanta GA 30332, USA}}\\
{\small {\it $^{+}$ Max-Planck-Institut f\"ur Astrophysik, 85748 Garching, Germany}}\\
\end{center}
\vskip 5mm


\begin{abstract}

The coalescence of a neutron-star binary is likely to result in the formation of a neutron-star merger remnant for a large range of binary mass configurations. The massive merger remnant shows strong oscillations, which are excited by the merging process, and emits gravitational waves. Here we discuss possibilities and prospects of inferring unknown stellar properties of neutron stars by the detection of postmerger gravitational-wave emission, which thus leads to constraints of the equation of state of high-density matter. In particular, the dominant oscillation frequency of the postmerger remnant provides tight limits to neutron-star radii. We mention first steps towards a practical implementation of future gravitational-wave searches for the postmerger emission. Moreover, we outline possibilities to estimate the unknown maximum mass of nonrotating neutron stars from such types of measurements. Finally, we review the origin and scientific implications of secondary peaks in the gravitational-wave spectrum of 
neutron-star mergers and differences in the dynamical behavior of the postmerger remnant depending on the binary configuration. These considerations lead to a unified picture of the post-merger gravitational-wave emission and the post-merger dynamics.

\end{abstract}


\section{Overview}
Neutron-star binaries are known to merge within the Hubble time for sufficiently close orbital separations because of the emission of gravitational waves, which reduce the angular momentum and energy of the system. Prior to merging the gravitational-wave emission is strongest, which is why the late phase of the so-called inspiral and the postmerger evolution are important targets for the upcoming gravitational-wave instruments (see e.g. Ref.~\cite{Faber2012}). Depending on the uncertain merger rate and the available detector network between 0.2 and 200 detections per year are expected~\cite{Abbott2013}.

The dynamics and the gravitational-wave signal of the inspiral phase are mostly determined by the masses of the binary, which is why a detection of gravitational waves from a neutron-star merger will yield the binary masses. In particular, the total binary mass can be well determined since it can be inferred from the accurately measured chirp mass (see Fig.~1 in Ref.~\cite{Bauswein2015epja}). In the very late inspiral phase finite-size effects may be measurable and may thus allow to infer unknown stellar properties of the inspiralling neutron stars like the stellar compactness, which constrains the only incompletely known high-density equation of state (e.g. Ref.~\cite{Read2013}). Complementary to that, the frequencies of the oscillations of the postmerger remnant yield information on stellar properties and the high-density equation of state. The postmerger oscillations are the focus of this study.

\section{Dominant postmerger oscillation}

\begin{figure}[h]
\begin{center}
\includegraphics[width=2.6in]{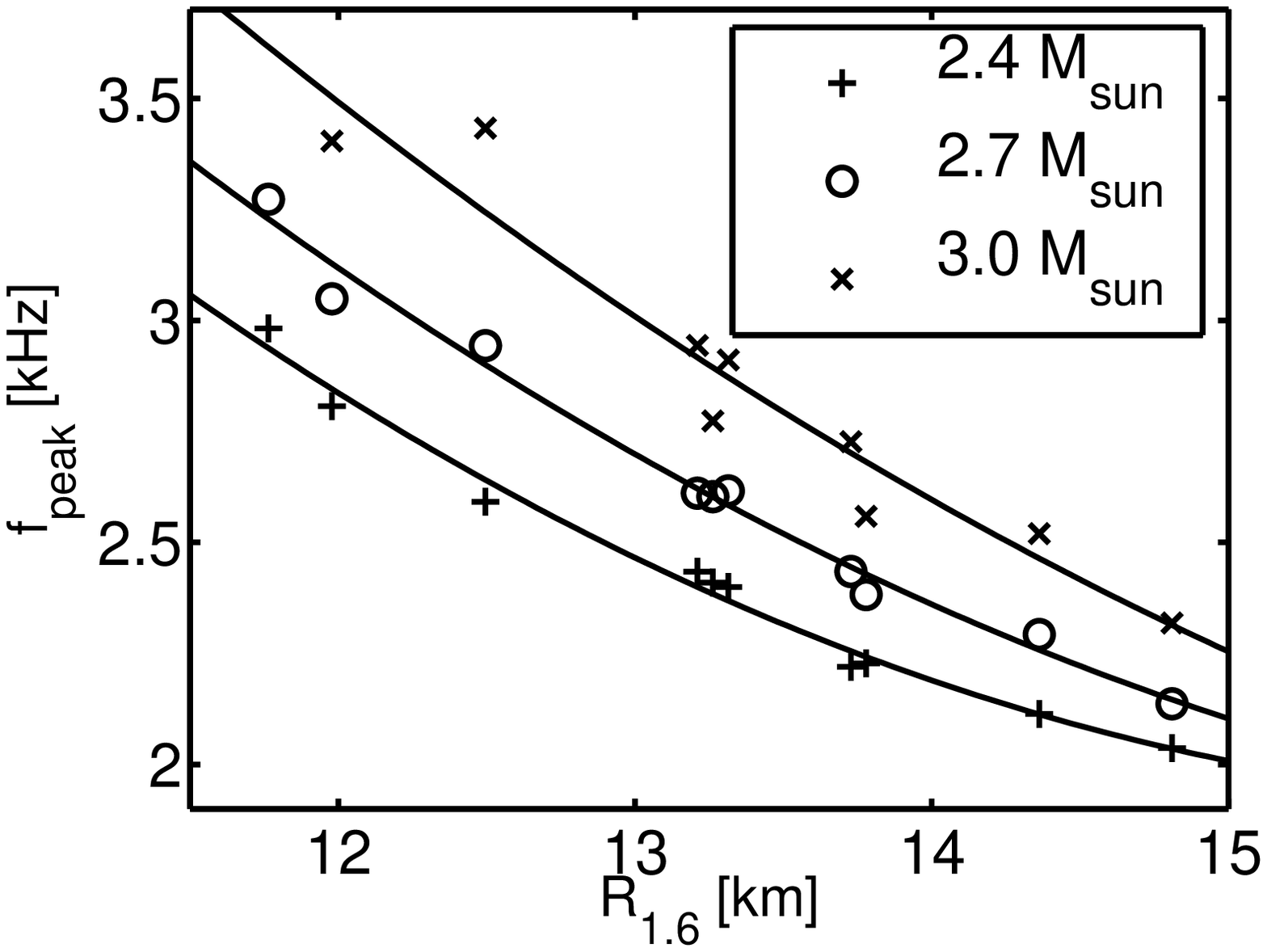}\includegraphics[width=2.6in]{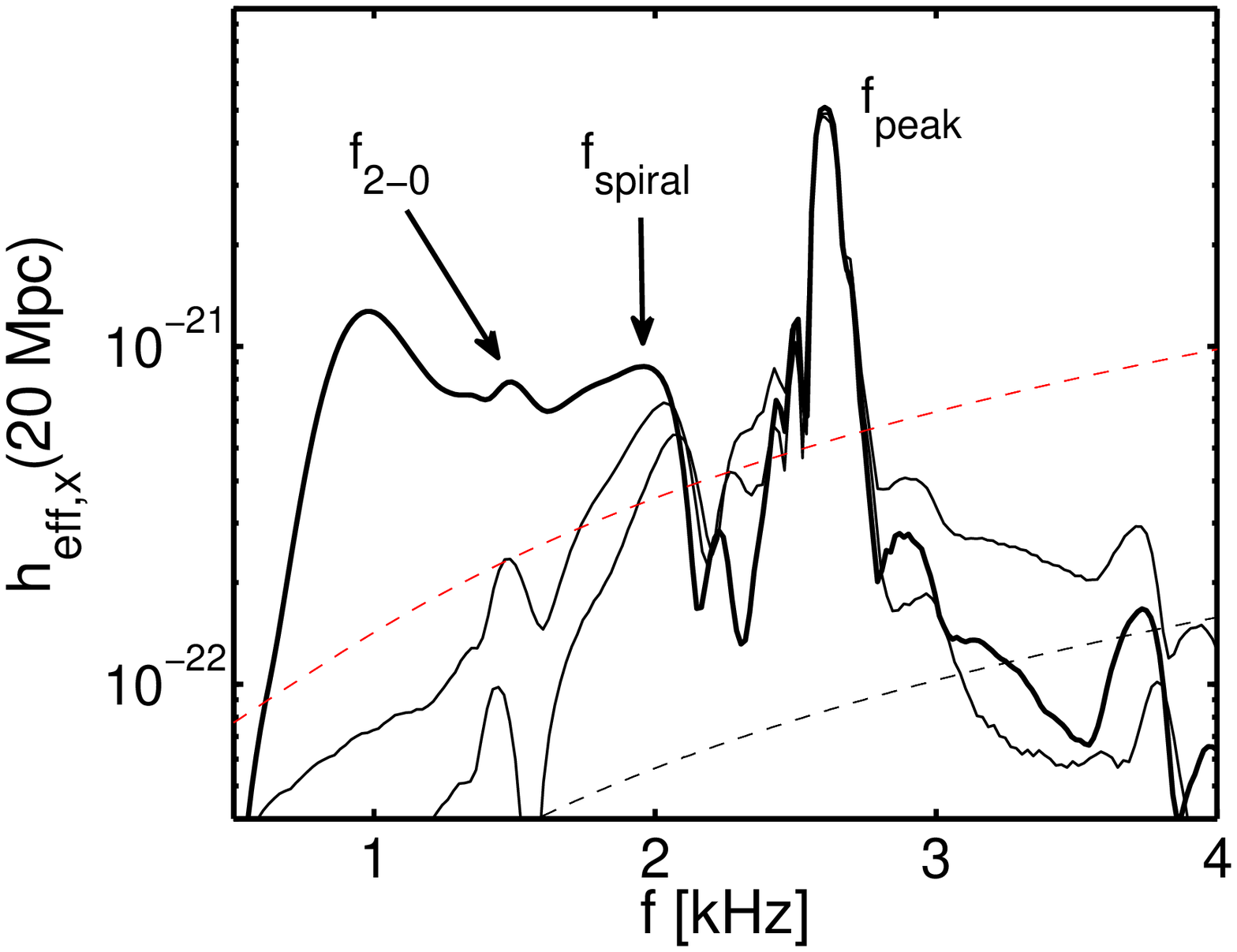}
\end{center}
\caption{Left panel: Relations between the dominant gravitational-wave frequency $f_\mathrm{peak}$ and the radii of nonrotating neutron stars for different fixed total binary masses. (Figure taken from Ref.~\cite{Bauswein2015epja}.) Right panel: Exemplary gravitational-wave spectrum with characteristic frequency peaks of the postmerger phase labelled by $f_{2-0}$, $f_\mathrm{spiral}$ and $f_\mathrm{peak}$. (Figure taken from Ref.~\cite{Bauswein2015epja}.)}
\label{fig1}
\end{figure}

For a large range of binary parameters the merging leads to the formation of a massive, differentially rotating neutron-star merger remnant~\cite{Bauswein2013}. The dynamics of the postmerger phase are strongly affected by the equation of state, which determines the structure of the merger remnant. Therefore, the dominant oscillation frequency of the remnant is characteristic of the high-density equation of state. Specifically, the dominant oscillation frequency scales tightly with the radii of nonrotating neutron stars of a fiducial mass~\cite{Bauswein2012prl,Bauswein2012prd}. Radii of nonrotating neutron stars depend on the equation of state and can thus be used to characterize a given equation of state. The fiducial mass can be chosen to be the one of the inspiralling stars. For a fixed binary mass the relation between the dominant frequency and the radius is even tighter if a fiducial mass somewhat higher than the mass of the inspiralling stars is chosen~\cite{Bauswein2012prd}. The reason is that the 
densities in the merger remnant are higher than in the initial stars. Hence, the stellar properties of more massive nonrotating stars provide a better description of equation-of-state properties in the density regime of the merger remnant. The relations between the dominant oscillation frequency $f_\mathrm{peak}$ and radii of non-rotating neutron stars, each radius value corresponding to a particular neutron-star equation of state, are shown in the left panel of Fig.~\ref{fig1} for different fixed binary masses.

The dominant oscillation frequency of the merger remnant occurs as a pronounced peak in the gravitational-wave spectrum (e.g. Refs.~\cite{Bauswein2012prl,Bauswein2012prd,Hotokezaka2013,Takami2014}), and there is strong evidence that this peak is generated by the fundamental quadrupolar fluid mode of the postmerger remnant~\cite{Stergioulas2011,Bauswein2015epja}. For an exemplary model the spectrum is provided in the right panel of Fig.~\ref{fig1} with the dominant oscillation frequency labelled by $f_\mathrm{peak}$. A measurement of this frequency yields tight constraints on neutron-star radii by inverting the empirical relation between the dominant postmerger oscillation frequency and the neutron-star radius (left panel of Fig.~\ref{fig1}). Figure~\ref{fig1} reveals that different relations exist for different total binary masses~\cite{Bauswein2012prl,Bauswein2012prd,Bauswein2015,Bauswein2015epja}. In a measurement the binary mass will be determined from the inspiral phase, and thus the relation to be 
employed for the 
frequency-radius inversion will be known.

The dominant peak in the gravitational-wave spectrum has been shown to be measurable by the upcoming network of gravitational-wave detectors for nearby merger events. Simulations reveal that a morphology-independent burst search algorithm can accurately determine the frequency of the most prominent peak in the kHz range of the gravitational-wave spectrum~\cite{Clark2014}. For distances within several to a few ten Mpc the dominant postmerger frequency can be measured with an accuracy of $\sim 10$~Hz. For merger rates on the more optimistic side, detections are possible with the upcomning network of instruments, and further improvements in the detectability are likely for more sophisticated search algorithms. For instance, a study based on a principal component analysis promises significant improvements in the detectability~\cite{Clark2015}.

In the case that several of such measurements succeed for different binary masses, one can consider the mass dependence of the peak frequency to estimate by an extrapolation the frequency at higher binary masses~\cite{Bauswein2014}. (The frequency increases with the binary mass since the remnant gets more compact with mass.) In this way the highest possible frequency and the highest possible binary mass which leads to a neutron-star merger remnant, can be approximately determined. Mergers with higher binary masses lead to a prompt gravitational collapse and directly form a black hole. The estimate of the threshold mass to prompt gravitational collapse can be employed to yield an estimate of the maximum mass of nonrotating neutron stars~\cite{Bauswein2013}. Similarly, the peak frequency close to the threshold determines the radius of the maximum-mass configuration of nonrotating neutron stars. Alternative approaches are summarized in Ref.~\cite{Bauswein2015epja}.

\section{Secondary postmerger frequencies}

Apart from the dominant oscillation frequency there are additional weaker features present in the gravitational-wave spectrum (right panel of Fig.~\ref{fig1}). Two distinct mechanisms can generate additional peaks depending on the binary mass~\cite{Bauswein2015prd}. Which of these mechanisms is active, depends on the binary mass. For relatively low masses antipodal bulges are formed during the merging, which orbit around the central parts of the remnant for a few milliseconds. The orbital motion of these bulges with a spiral pattern generates a pronounced secondary peak several 100 Hz below the dominant oscillation frequency ($f_\mathrm{spiral}$ in Fig.~\ref{fig1})~\cite{Bauswein2015prd}. As the binary mass increases, the initial binary components are more compact and merge with a higher impact velocity. The more violent collision leads to a stronger excitation of the quasi-radial oscillation mode of the merger remnant, which typically has a frequency of roughly 1000 Hz. While this mode being quasi-radial 
does not appear in the gravitational-wave spectrum, it can couple to the dominant 
quadrupolar mode discussed above. The interaction of these two modes leads to two distinct peaks at frequencies of the dominant mode shifted by $\pm$ the frequency of the radial mode~\cite{Stergioulas2011}. In particular, the peak at lower frequencies can be very pronounced, which occurs then at about 1000 Hz below the dominant peak ($f_{2-0}$ in Fig.~\ref{fig1}). For moderate binary masses the two mechanisms can lead to peaks with comparable strength in the gravitational-wave spectrum like in the case shown in Fig.~\ref{fig1}. As the binary mass approaches the threshold binary mass to the prompt gravitational collapse from below, the quasi-radial oscillations become stronger. Hence, the gravitational-wave peak generated by the mode coupling starts to dominate over the peak produced by the spiral pattern. The formation of antipodal bulges is weakened for more compact binary components. The picture of distinct mechanisms producing peaks in the gravitational-wave spectrum at different frequencies is supported 
by the analysis of time-frequency maps~\cite{Clark2015}.

\begin{figure}[h]
\begin{center}
\includegraphics[width=2.6in]{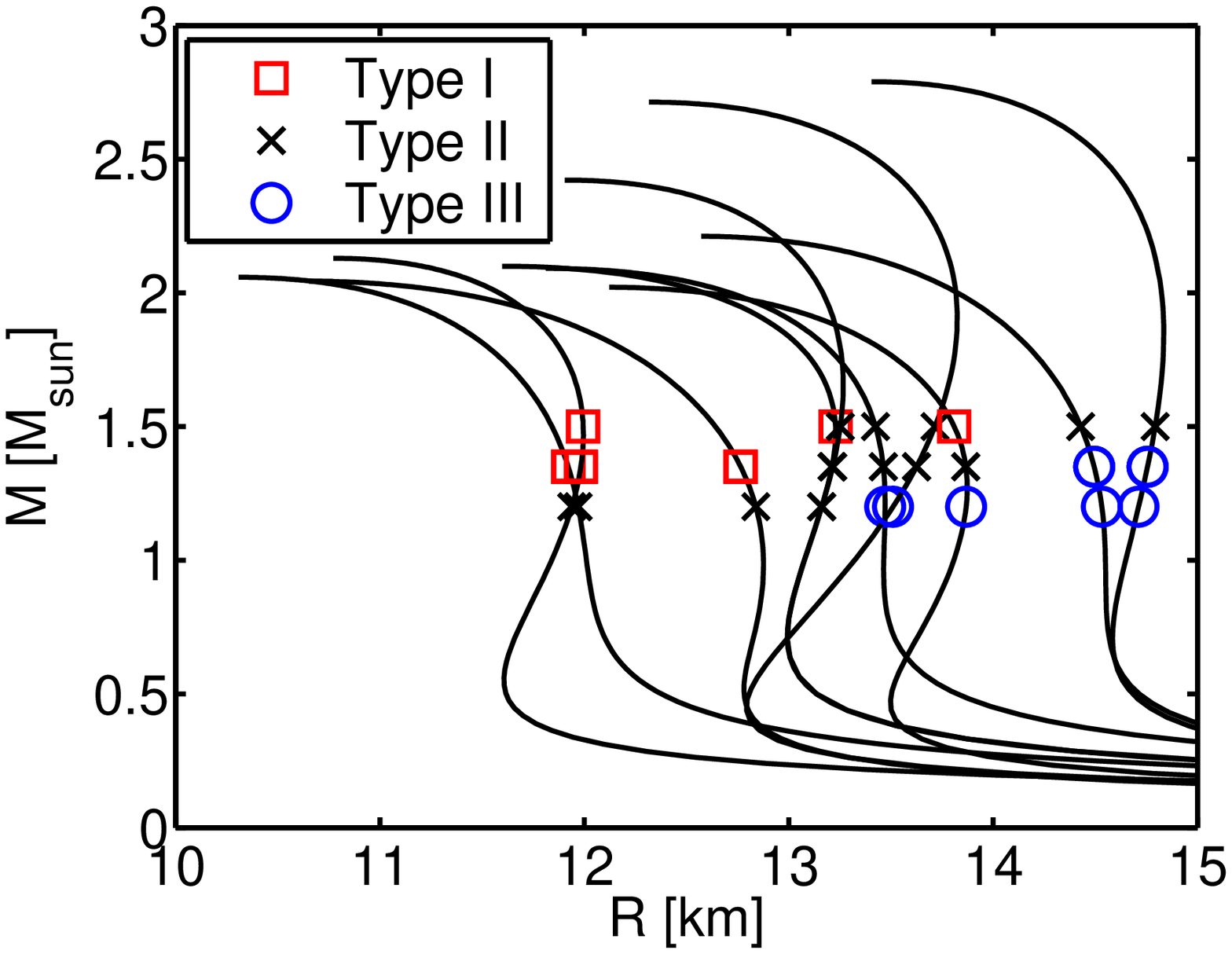}\includegraphics[width=2.6in]{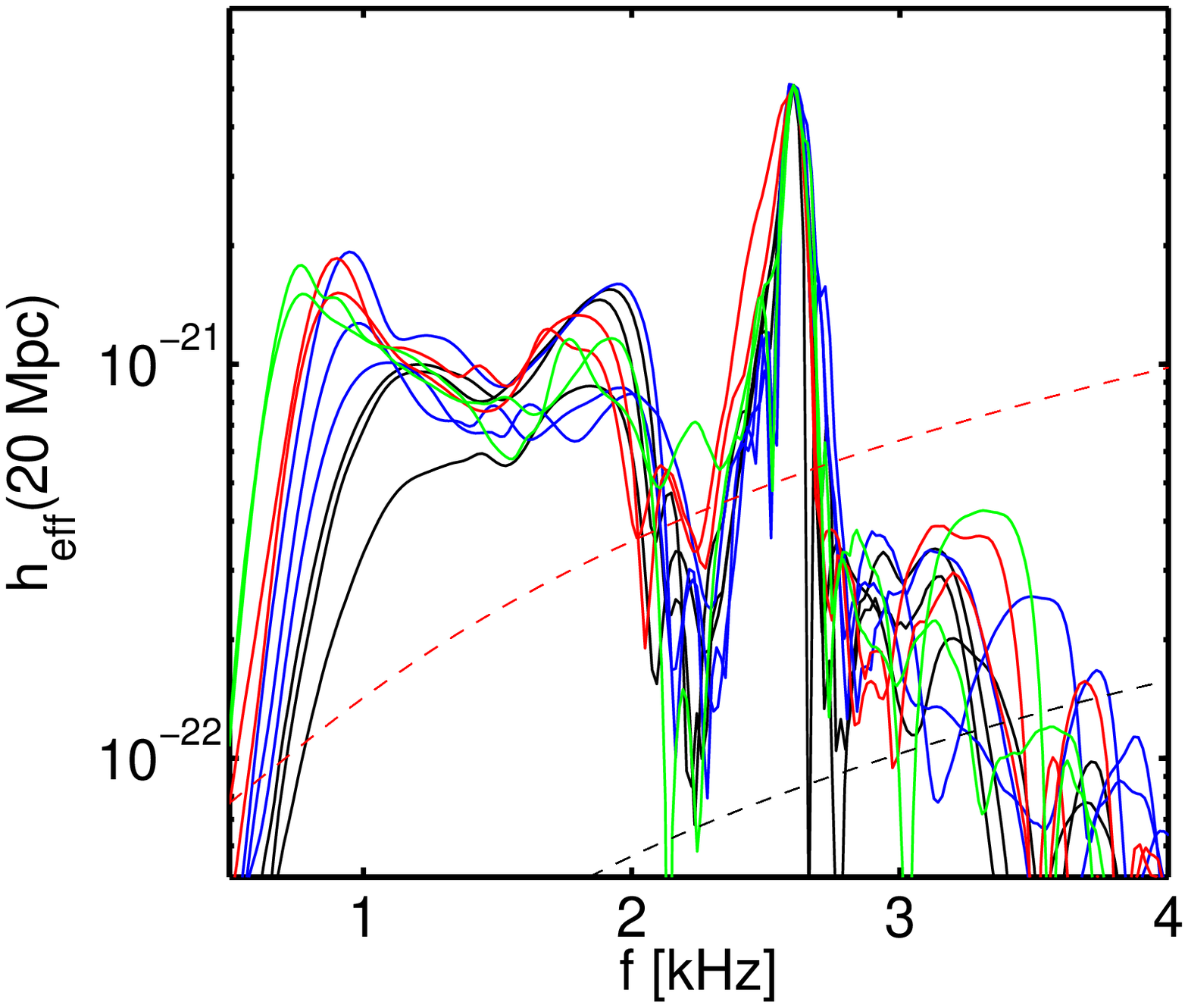}
\end{center}
\caption{Left panel: Classification scheme of postmerger dynamics and gravitational-wave emission (Type I: dominant $f_{2-0}$ feature, Type II: comparable strength of $f_{2-0}$ and $f_\mathrm{spiral}$ feature, Type III: dominant $f_\mathrm{spiral}$ feature). Shown are the types of gravitational-wave spectra for different equations of state and total binary masses $M_\mathrm{tot}$. The type of a given model is plotted at $M_\mathrm{tot}/2$ (and the radius of the inspiralling neutron stars) on top of the mass-radius relations of the equation of state which was used in the hydrodynamical simulation (see Ref.~\cite{Bauswein2015prd} for details). (Figure taken from Ref.~\cite{Bauswein2015prd}.) Right panel: Gravitational-wave spectra for different equations of state and fixed total binary mass rescaled to a reference frequency and amplitude.}
\label{fig2}
\end{figure}

In these considerations the notion of high and low total binary masses depends on the equation of state because the equation of state determines the stellar compactness and the threshold to the prompt gravitational collapse. A detailed analysis of many different models (sampling different binary masses and equations of state) leads to a classification scheme of the postmerger dynamics and gravitational-wave emission. The scheme classifies gravitational-wave spectra depending on the presence and strength of the different secondary peaks (see left panel of Fig.~\ref{fig2}). It illustrates the aforementioned binary mass dependence and also reveals the clear impact of the equation of state consistent with the picture described above~\cite{Bauswein2015prd}. For stiff equations of state the generation of the $f_\mathrm{spiral}$ feature is favored, which leads to a strong secondary peak even for relatively high total binary masses. In the case of soft equations of state, already for relatively low total binary 
masses the spiral feature is suppressed, while the radial mode and the corresponding secondary peak are strongly present. This explains the band structure in Fig.~\ref{fig2}.

Remnants with higher mass, which are closer to the gravitational collapse, have a lower quasi-radial frequency, which should go to zero for the onset of the collapse. Consequently, the secondary 
peak produced by the mode coupling gets closer to the dominant peak. A determination of the frequency of this secondary peak and the main peak determines the quasi-radial frequency of the remnant and may indicate the proximity of the system to the quasi-radial gravitational collapse, which may possibly yield an estimate of the maximum mass of nonrotating neutron stars~\cite{Bauswein2015prd}.

It is interesting to note that also the frequencies of the different secondary peaks depend in a clear way on the equation of state~\cite{Bauswein2015prd}. In similarity to the dependence of the dominant frequency $f_\mathrm{peak}$ discussed above, the secondary frequencies also scale tightly with stellar radii of nonrotating neutron stars for fixed total binary masses. This behavior may be expected considering the mechanisms that generate those secondary peaks. Equivalently, one finds also a strong correlation between the frequencies of the secondary frequency and the dominant frequency for a fixed total binary mass (see Fig.~4 in Ref.~\cite{Clark2015}). The relation can be well described by a linear function $f_\mathrm{spiral\, / \, 2-0}\propto f_\mathrm{peak}$. The importance of this observaiton lies in the fact that it explains the existence of a universal gravitational-wave spectrum, which can be obtained by rescaling the frequency and amplitude of a given spectrum to a reference point~\cite{Clark2015}.
 The 
right panel of Fig.~\ref{fig2} demonstrates the universality, i.e. the similarity of the rescaled spectra and their approximate independence of the equation of state. The approximate universality of postmerger gravitational-wave spectra may be regarded as the reason why the principal component analysis presented in Ref.~\cite{Clark2015} performs so well.

The understanding of the different mechanisms, which generate the most prominent peaks in the postmerger gravitational-wave spectrum, leads to a relatively simple, physically motivated time-domain model of the gravitational-wave signal. As shown in Ref.~\cite{Bauswein2015epja} (Fig.~12 to~14), the sum of damped sine functions performs very well in reproducing the gravitational-wave signals which are obtained from hydrodynamical simulations. The observed match between our analytic model and the numerical data is very encouraging and our model may form the basis for future time-domain templates to be used in matched-filtering searches of gravitational waves. The performance of this model still needs to be confirmed for a wider range of numerical models.

\section*{Acknowledgements}
A.B. is a Marie Curie Intra-European Fellow within the 7th European Community Framework Programme (IEF 331873). A.B. acknowledges support by the Klaus Tschira Foundation.


\vskip 10mm

\begin{thebibliography}{99}

\bibitem{Faber2012}
J.~A. {Faber} and F.~A. {Rasio}, {\it Liv.
  Reviews in Relativity} {\bf 15}, 8 (2012).


\bibitem{Abbott2013}
  The LIGO Scientific Collaboration and the Virgo Collaboration and B.~P. {Abbott} et al., {\it ArXiv:1304.0670}.
  
\bibitem{Bauswein2015epja}
  A.~{Bauswein}, N.~{Stergioulas} and H.-T. {Janka}, {\it accepted for publication in Eur. Phys. J. A (2015), ArXiv:1508.05493}.
  
\bibitem{Read2013}
J.~S. {Read}, L.~{Baiotti}, J.~D.~E. {Creighton}, J.~L. {Friedman},
  B.~{Giacomazzo}, K.~{Kyutoku}, C.~{Markakis}, L.~{Rezzolla}, M.~{Shibata} and
  K.~{Taniguchi}, {\it Phys. Rev. D} {\bf 88}, 044042 (2013).

\bibitem{Bauswein2013}
A.~{Bauswein}, T.~W. {Baumgarte} and H.-T. {Janka}, {\it Phys. Rev. Lett.} {\bf 111},
  131101 (2013).
  
\bibitem{Bauswein2012prl}
A.~{Bauswein} and H.-T. {Janka}, {\it Phys. Rev. Lett.} {\bf 108}, 011101 (2012).

\bibitem{Bauswein2012prd}
A.~{Bauswein}, H.-T. {Janka}, K.~{Hebeler} and A.~{Schwenk}, {\it Phys. Rev. D} {\bf 86}, 063001 (2012).

\bibitem{Hotokezaka2013}
K.~{Hotokezaka}, K.~{Kiuchi}, K.~{Kyutoku}, T.~{Muranushi}, Y.-i. {Sekiguchi},
  M.~{Shibata} and K.~{Taniguchi}, {\it
  Phys. Rev. D} {\bf 88}, 044026 (2013).

\bibitem{Takami2014}
K.~{Takami}, L.~{Rezzolla} and L.~{Baiotti}, {\it Phys. Rev. Lett.}
  {\bf 113}, 091104 (2014).

\bibitem{Stergioulas2011}
N.~{Stergioulas}, A.~{Bauswein}, K.~{Zagkouris} and H.-T. {Janka},
   {\it Mon. Not. R. Astron. Soc.} {\bf 418}, 427 (2011).
  
\bibitem{Bauswein2015}
A.~{Bauswein}, N.~{Stergioulas} and H.-T. {Janka}, {\it ArXiv:1503.08769}.
  
\bibitem{Clark2014}
J.~{Clark}, A.~{Bauswein}, L.~{Cadonati}, H.-T. {Janka}, C.~{Pankow} and
  N.~{Stergioulas}, {\it Phys. Rev. D} {\bf 90}, 062004 (2014).

\bibitem{Clark2015}
  J.~{Clark}, A.~{Bauswein}, N.~{Stergioulas} and D.~{Shoemaker}, {\it accepted for publication in Class. Quantum Gravity (2015), ArXiv:1509.08522}.
  
\bibitem{Bauswein2014}
A.~{Bauswein}, N.~{Stergioulas} and H.-T. {Janka}, {\it Phys. Rev. D} {\bf 90}, 023002 (2014).

\bibitem{Bauswein2015prd}  
  A.~{Bauswein} and N.~{Stergioulas}, {\it Phys. Rev. D} {\bf 91}, 124056 (2015).



\end{thebibliography}
\end{document}